\documentclass[doublecol,figures]{epl2}
\usepackage{amssymb,amsmath}
\usepackage{epsf,colordvi,graphicx,color,amsbsy}
\usepackage[T1]{fontenc}

\title{Capillary wave turbulence on a spherical  fluid surface\\ in low gravity}
\author{C. Falc\'on\inst{1} \and E. Falcon\inst{2}\thanks{Corresponding author: \email{eric.falcon@univ-paris-diderot.fr}} \and U. Bortolozzo\inst{1,2} \and S. Fauve\inst{1}}
\shortauthor{C. Falc\'on \etal}
\institute{                    
\inst{1} Laboratoire de Physique Statistique (LPS), \'Ecole Normale Sup\'erieure, CNRS -- UMR 8550\\ 24, rue Lhomond, 75 005 Paris, France, EU\\
\inst{2} Laboratoire Mati\`ere et Syst\`emes Complexes (MSC), Universit\'e Paris Diderot, CNRS -- UMR 7057\\ 10 rue A. Domon \& L. Duquet, 75 013 Paris, France, EU
}

\pacs{47.35.Pq}{Capillary waves} 
\pacs{47.52.+j}{Chaos in fluid dynamics} 
\pacs{47.54.Ðr}{Pattern selection; pattern formation} 
\pacs{81.70.Ha}{Testing in microgravity environments} 

\abstract{We report the observation of capillary wave turbulence on the surface of a fluid layer in a low-gravity environment. In such conditions, the fluid covers all the internal surface of the spherical container which is submitted to random forcing. The surface wave amplitude displays power-law spectrum over two decades in frequency, corresponding to wavelength from $mm$ to a few $cm$. This spectrum is found in roughly good agreement with wave turbulence theory. Such a large scale observation without gravity waves has never been reached during ground experiments.  When the forcing is periodic, two-dimensional spherical patterns are observed on the fluid surface such as subharmonic stripes or hexagons with wavelength satisfying the capillary wave dispersion relation.}

\begin{document}

\maketitle

Wave turbulence concerns the study of the dynamical and statistical properties of an ensemble of dispersive waves with nonlinear interactions.  Wave turbulence occurs at very different scales in a great variety of systems: surface or internal waves in oceanography \cite{Toba73,Lvov04}, Alfv\'en waves in solar wind \cite{Sagdeev79}, plasmas \cite{Mizuno83}, surface waves on elastic plates \cite{Rica}, spin waves in solids. Surprisingly, only a few groups have performed laboratory experiments on this subject so far, mainly focusing on wave turbulence on a fluid surface \cite{Wright96,Lommer02,Brazhnikov02,Onorato04,Falcon07}. These wave turbulence experiments are scarce compared to numerous studies in hydrodynamic turbulence, although various analytical results have been obtained in the framework of wave turbulence or ``weak turbulence'' theory \cite{Zakharov}.   

Gravity and capillary turbulent wave regimes on a fluid surface are characterized by different Kolmogorov type spectra \cite{Falcon07}. These two regimes influence each other and coexist at different scales in the same experiment. Since energy tranfers in capillary and gravity regimes are not governed by similar nonlinear processes, it is of primary interest to study a pure capillary wave regime. The importance of the gravity and the capillary effects is quantified by the ratio between the wavelength of the surface wave, $\lambda$, and the capillary length, $l_c \equiv \sqrt{\gamma/(\rho g)}$, where $\gamma$ and $\rho$ are the surface tension and the density of the fluid, respectively, and $g$ is the acceleration of gravity. For usual fluids, $l_c$ is of the order of few mm, corresponding to a critical wavelength $\lambda_c=2\pi l_c$ of the order of 1 cm. Gravity waves are thus proeminent for wavelength larger than few cm. The capillary length cannot be significantly changed using other interfaces between simple fluids and air. It is in an intermediate range between the size of the experiment and the dissipative length. In usual laboratory-scale experiments, this limits both the gravity and capillary regimes to less than a decade in frequency. In a low-gravity experiment, one can obtain capillary waves at all wavelengths of the fluid container of size $L$, provided that $\lambda_c>L$.
We emphasize that increasing the capillary range toward large frequencies using better resolved measurements in ground experiments, without suppressing the development of gravity waves, is not satisfactory. Gravity and capillary waves indeed interact because the energy flux cascades among the different scales. In addition, it is predicted that weak turbulence for capillary waves breaks down at large scale, because the nonlinear time scale becomes comparable to the linear one~\cite{Connaughton03}. It is thus of interest to remove gravity waves that may hide this phenomenon.

Here, we report the observation of the power spectrum density of the capillary wave turbulence regime over a large range of wavelengths in a low-gravity environment. The invariant-scale power spectrum is found in roughly good agreement with weak turbulence theory.  We also study parametric excitation of a fluid in low gravity by sinusoidally forcing its container. Although the so-called ``Faraday instability'' has been extensively studied with gravity \cite{Faraday31,Benjamin54,Edwards94}, only one trial has been performed in low gravity near the critical point \cite{Beysens98}. We report here the first experimental observation of two-dimensional wave patterns on a spherical or cylindrical fluid surface in low gravity. Applications of this work could be extended to the lattice wrapping on the curved surfaces as well as in condensed matter such as in spherical crystallography \cite{Bausch03}. In addition, pattern formation in spherical geometry for axisymmetric systems is of obvious interest in the context of geophysical and astrophysical fluid dynamics~\cite{futterer}. 
%For instance, a model of convection in the Earth core will be studied in microgravity using two co-rotating concentric spheres with a central force field generated by applying a high voltage difference~\cite{futterer}. Parametric forcing in spherical geometry has been studied in the context of  pulsating stars~\cite{poyet}.

\begin{figure}[t]
\centerline{
\epsfxsize=82mm
\epsffile{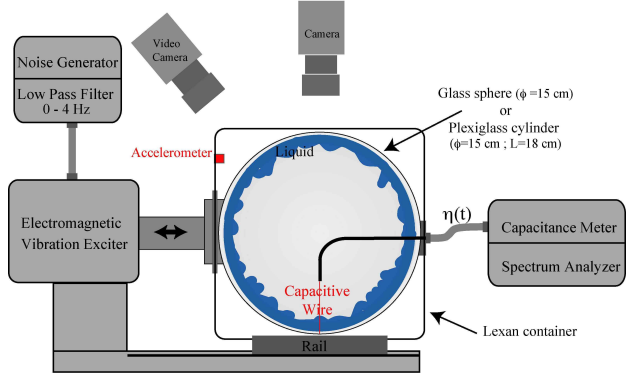} 
}
\caption{Sketch of the experimental setup. In microgravity phases, the fluid covers all the internal surface of the container submitted to vibrations.}
\label{fig01}
\end{figure}

\begin{figure}[t]
\centerline{
%\begin{tabular}{c}
\epsfxsize=65mm \epsffile{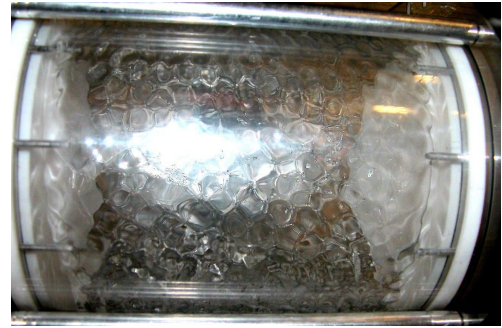} %Vol3Cy1527.eps
%\\
%\epsfxsize=65mm \epsffile{stripes2.eps} %Vol3Cy1532.eps
%\end{tabular}
}
\caption{Two-dimensional cylindrical subharmonic wave patterns (hexagons) under a sinusoidal forcing at frequency $f_0=30$ Hz, and amplitude $d_0=0.29$ mm leading to a 1.06g acceleration. Cylindrical container filled with 30 cl of ethanol.}
%a) stripes ($f_0=30$ Hz) 
\label{fig02}
\end{figure}
\begin{figure}[Ht!]
\centerline{
%\begin{tabular}{cc}
%\epsfxsize=50mm \epsffile{Vol3Sp1560ter.eps} 
%\\
\epsfxsize=60mm \epsffile{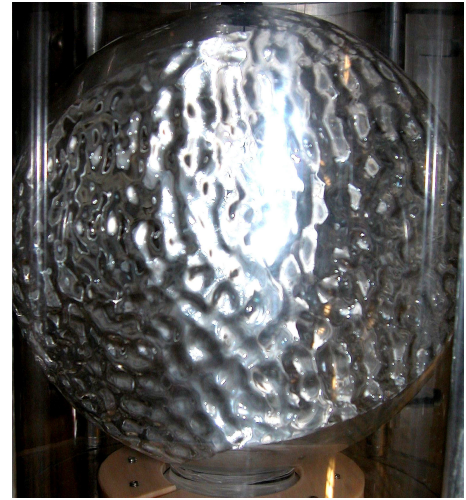} %Vol1SP1497ter.eps
%\end{tabular}
}
\caption{Two-dimensional spherical subharmonic wave patterns under sinusoidal forcing at frequency $f_0=30$ Hz, and amplitude $d_0=0.32$ mm leading to a 1.16g acceleration. Spherical container filled with 20 cl of water.}
\label{fig03}
\end{figure}

The experimental setup is sketched in Fig.~\ref{fig01}. A container partially filled with a fluid is put down on a rail, and is submitted to vibrations by means of an electromagnetic exciter (BK~ 4803) via a power amplifier (BK~2706). To study wave turbulence, the container is driven with a random forcing, supplied by the source of a dynamical analyzer (Agilent~35 670A), and low-pass filtered in the frequency range 0 - 6 Hz. This corresponds to wavelengths of surface waves larger than 1 cm in low gravity. To study wave patterns, the container is driven with a sinusoidal forcing at frequency $f_0$ in the range $10\leq f_0 \leq 70$ Hz, and amplitude $d_0 \sim $ mm corresponding to a container acceleration $0.1 \lesssim a_0 \lesssim 30$g. The container geometry is either spherical (15 cm in diameter) or cylindrical (15 cm in diameter, 18 cm in length). Each container is made of a wetting material (Plexiglas cylinder or glass sphere) to avoid that the fluid loses contact with the internal wall of the container in low-gravity environment. According to its geometry, the container is filled with 20 or 30 cl of fluid. This corresponds to an uniform fluid layer of roughly 5 mm depth covering all the internal surface of the container in low-gravity environment. The fluid is either ethanol or water.
The local displacement of the fluid is measured with a capacitive wire gauge, 0.1 mm in diameter, plunging into the fluid \cite{Falcon07}. This sensor allows wave height measurements from 10 $\mu$m up to 2 cm with a 0.1 ms response time. A piezoelectric accelerometer (PCB)  is screwed on the container to record its acceleration. A dynamical analyzer is used to record the power spectrum of the surface wave amplitude during each microgravity phase. The motion of the fluid surface is visualized with a Nikon camera and recorded with a Sony video camera. Microgravity environment  (about $\pm 5\times 10^{-2}$g) is repetitively achieved by flying with the specially modified {\em Airbus A300 Zero-G} aircraft through a series of parabolic trajectories which result in low-gravity periods, each of 22 s. We observe that the fluid crawls up the sides of the container and covers all the internal surface of the tank due to the capillary forces. This takes roughly 1 s. Measurements have thus been recorded only on $18$ s to eliminate these transients.
Contrary to the common sense, no formation of a single sphere of fluid is observed in the middle of the tank, due to these capillary effects. An homogeneous fluid layer is formed on the internal surface of the tank, confining air in its center. 
When the container is submitted to a sinusoidal forcing at frequency $f_0$, surface wave patterns appear as shown in Fig.\ \ref{fig02} with a cylindrical container. These two-dimensional patterns are either stripes (not shown here) or hexagons (see Fig.\ \ref{fig02}) depending on the vibrating frequency. By using another container geometry, one can also observe spherical patterns as shown in Fig.\ \ref{fig03}. 

\begin{figure}[t]
\centerline{
\epsfysize=60mm
\epsffile{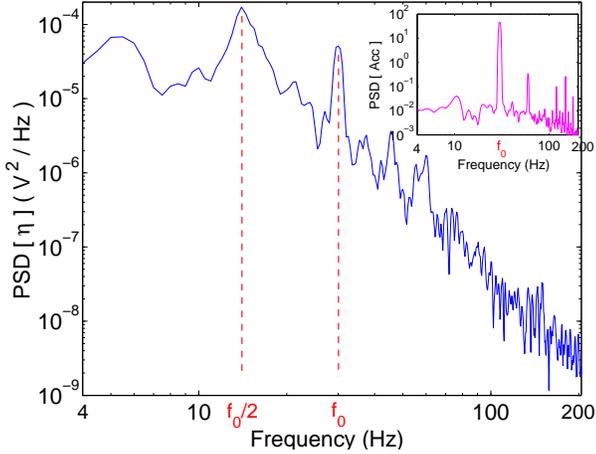} 
}
\caption{Typical subharmonic response of patterns in low-gravity. Power spectrum density of surface wave height. $f_0=30$ Hz is the driving frequency. $f_0/2$ is the main response frequency. Inset: Power spectrum density (PSD) of the container acceleration showing the driving frequency $f_0$.}
\label{fig04}
\end{figure}

To understand the mechanism of pattern formation, one records simultaneously the acceleration imposed to the container, and the surface wave height as a function of time. A typical power spectrum density of the container acceleration is shown in the inset of Fig.\ \ref{fig04}. The main peak at $f_0=30$ Hz corresponds to the driving frequency. The typical response of the fluid surface to this excitation is displayed in Fig.\ \ref{fig04}. The power spectrum of surface waves shows two main peaks: a subharmonic one close to $f_0/2\simeq 15$ Hz, and a smaller one at $f_0$ corresponding to a  reminiscence of the driving frequency. The two-dimensional patterns are thus subharmonic ones. The pattern formation can be understood at first sight as simple parametric excitation in low gravity. However, the patterns are not stationary and their dynamics appears to be very complex: a sloshing motion which depends on the jitters of residual gravity is usually superimposed to the parametric excitation. A complete dynamical description of the pattern deserves further studies. 

\begin{figure}[t]
\centerline{
\epsfysize=60mm
\epsffile{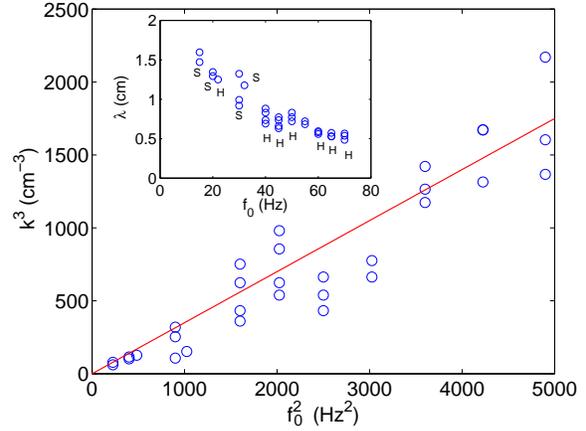} 
}
\caption{Pattern wave number cubed, $k^3$, as a function of the driving frequency squared, $f_0^2$. Solid line of slope $0.35$ s$^2$/cm$^3$. Inset: pattern wavelength, $\lambda$, as a function of the driving frequency, $f_0$. Symbols ``S'' and ``H'' correspond to stripe and hexagon patterns respectively. Fluid is ethanol.}
\label{fig05}
\end{figure}

The wavelength, $\lambda$, is then measured as a function of the driving frequency, $f_0$. As shown in the inset of Fig.\ \ref{fig05}, $\lambda$ is found to decrease with increasing $f_0$. Using the dispersion relation of pure capillary waves in the deep layer limit, $\omega^2=(\gamma/\rho)k^3$ with $k\equiv 2\pi/\lambda$ and $\omega=2\pi f$ where $f=f_0/2$ is the pattern frequency, one has $k^3=cf_0^2$ where $c=\pi^2\rho/\gamma$ is a constant depending on the fluid density, $\rho$, and surface tension,  $\gamma$. $k^3$ is indeed found roughly proportional to $f_0^2$ with the expected coefficient $c=\pi^2\rho/\gamma\simeq 0.35$ s$^2$/cm$^3$ (extracted from ethanol properties $\rho=790$ kg/m$^3$; $\gamma=22\times 10^{-3}$ N/m) as shown by the solid line in Fig.\ \ref{fig05}. The dispersion relation of capillary waves in low gravity is thus obtained even in the low frequency regime where gravity waves are usually present.

\begin{figure}[Ht!]
\centerline{
\epsfysize=60mm
\epsffile{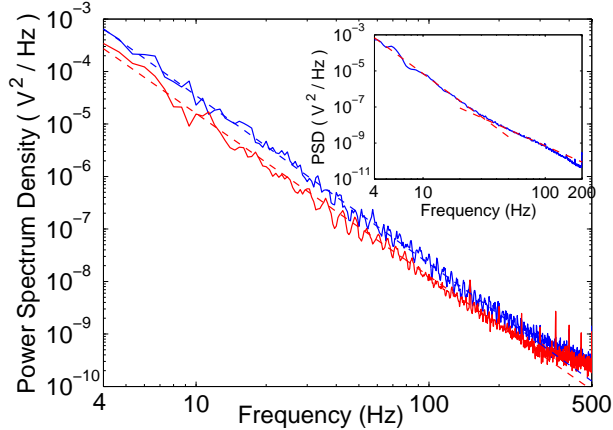}
}
\caption{Power spectrum density of surface wave height {\em in low gravity}. Lower curve: Random forcing 0 - 6 Hz. Upper curve: Sinusoidal forcing at 3 Hz. Dashed lines had slopes of -3.1 (lower) and -3.2 (upper). Cylindrical container filled with 30 cl of ethanol. Inset: same {\em with gravity}.  Slopes of dashed lines are -5 (upper) and -3 (lower) corresponding respectively to gravity and capillary wave turbulence regimes. Rectangular container filled with a 20 mm ethanol depth.}
\label{fig06}
\end{figure}

Let us now present our wave turbulence results in low gravity. To wit the cylindrical container is submitted to low-frequency random forcing in low-gravity. Surface waves with erratic motion then appear on the free surface as schematically shown in Fig.\ \ref{fig01}.  The power spectrum of surface wave amplitude then is recorded, and is shown in Fig.~\ref{fig06}. One single power-law spectrum is observed on two decades in frequency. Whatever the geometry of the tank (sphere or cylinder) and the large scale forcing (random or sinusoidal), the exponent is close to $-3$ (see Fig.~\ref{fig06}). This power-law exponent has roughly the same value that the one found under gravity for the capillary wave turbulence regime (see the inset of Fig.\ \ref{fig06}). Weak turbulence theory predicts a $f^{-17/6}$ scaling of the surface height spectrum for the pure capillary regime \cite{Zakharov67Cap}. This $-17/6 \simeq -2.8$  expected exponent  is close to the value -3 reported here. Kolmogorov-like spectrum of the capillary wave turbulence regime is thus observed in Fig.~\ref{fig06} over two decades in frequency. To our knowledge, this large range of frequencies has never been reached with ground experiments. This allows to study better the capillary wave turbulence regime. The power spectrum of surface height in the presence of gravity is shown for comparison in the inset of Fig.~\ref{fig06}. It displays two power laws:  $f^{-5}$ and $f^{-3}$ corresponding respectively to gravity and capillary wave turbulence regimes. For the gravity regime, the power law exponent is forcing dependent \cite{Falcon07,Denissenko07} in contrast with numerical \cite{Onorato02} and theoretical \cite{Zakharov} predictions in $f^{-4}$.
The capillary range is thus limited at low frequencies $f \leq f_c =\sqrt{g/(2 \pi^2l_c)}\propto g^{3/4} \sim 20$ Hz. The capillary length $l_c$ being of order of few mm for usual fluids, the critical frequency $f_c$ is in rough agreement with the one observed in the inset of Fig.~\ref{fig06} (see also Ref. \cite{Falcon07}). Such a critical frequency corresponds to a wavelength of the order of 1 cm. When $g \rightarrow 0$, the cross-over frequency between both regimes is then predicted to be pushed away to very low frequency. For our microgravity precision, $\pm 0.05g$, the capillary length then is expected to be close to cm, and the cross-over frequency of the order of 1 Hz, corresponding to wavelength of the order of 10 cm. Thus, in microgravity, for our frequency range (4 Hz up to 500 Hz), the power spectrum of surface wave amplitude is not polluted by gravity waves.
At high frequency,  the power spectrum in the capillary range in microgravity (Fig.~\ref{fig06}) is limited at frequency about $400$ Hz due to the low signal-to-noise ratio. Note that the high frequency limitation is lower in the presence of gravity ($\geq 100$ Hz) (see the inset of Fig.~\ref{fig06}). This cut-off frequency is related to the meniscus diameter on the capacitive wire gauge that prevents the detection of waves with a smaller wavelength. In microgravity, this latter effect vanishes since the meniscus diameter becomes of the order of the size of the container. 

We have reported the observation of capillary wave turbulence on a fluid surface in low-gravity. When the container is submitted to random forcing, we observe an invariant-scale power spectrum of wave amplitude on two decades in frequency in roughly good agreement with wave turbulence theory. This spectrum is independent on the large-scale forcing parameter. When the container is submitted to periodic forcing, we report the first observation of two-dimensional subharmonic patterns (stripes, hexagons) on a spherical or cylindrical fluid surface. Their wavelengths lead to a measurement of the dispersion relation of linear capillary waves in low gravity. These patterns results from a simple parametric excitation with no boundary effects. Their dynamical description is much more complex and results from the interaction between two instabilities (sloshing motion and parametric amplification). Note that the slope of the continuous part of the spectrum is steeper in the presence of parametric wave patterns than for wave turbulence ($f^{-4}$ in Fig.~\ref{fig04} instead of $f^{-3}$ in Fig.~\ref{fig06}).  This can be related to cusps of the spatial patterns sweeping the sensor~\cite{Kuznetsov04}. Understanding the differences between disordered parametric wave patterns and weak wave turbulence deserves further studies, in particular the simultaneous measurement of temporal and spatial spectra~\cite{Savelsberg}.
%effect of the spatial pattern sweeping the measurement wire.

\acknowledgments
We greatly acknowledge Y. Garrabos and the Novespace team for their technical assistance. This work has been supported by the CNES and by ANR turbonde BLAN07-3-197846. The flight has been provided by Novespace. {\em Airbus A300 Z\'ero-G} aircraft is a program of CNES. C.~F. has a fellowship of CONYCIT and U.~B. a fellowship of Ville de Paris and of the European Commission (MEIF-CT-2006-041594).

%%%%%%%%%%%%%%%%%%%%%%%%%%%%%%%%%%%%%%
%%%%%%%%%%%% REFERENCES %%%%%%%%%%%%%%%%%%
%%%%%%%%%%%%%%%%%%%%%%%%%%%%%%%%%%%%%%

\end{document}